\documentclass[aps,pre,twocolumn,groupedaddress,superscriptaddress,showpacs]{revtex4-1}
\usepackage{epsfig,amssymb,amsmath,graphicx,subfigure,hyperref}
\usepackage{tabularx}
\usepackage{float}
\usepackage{setspace}
\usepackage{color}

\usepackage{array} 
\newcommand{\PreserveBackslash}[1]{\let\temp=\\#1\let\\=\temp} \newcolumntype{C}[1]{>{\PreserveBackslash\centering}p{#1}} \newcolumntype{R}[1]{>{\PreserveBackslash\raggedleft}p{#1}} \newcolumntype{L}[1]{>{\PreserveBackslash\raggedright}p{#1}}

\begin{document}

\title{Thermal stiffening of clamped elastic ribbons}

\author{Duanduan Wan}
\altaffiliation{Current address: Department of Chemical Engineering, University of Michigan, Ann Arbor, Michigan 48109, USA}
\affiliation{Soft Matter Program and Department of Physics, Syracuse University, Syracuse, New York 13244-1130, USA}
\author{David R. Nelson}
\affiliation{Departments of Physics, and Molecular and Cellular Biology and School of Engineering and Applied Sciences, Harvard University, Cambridge, Massachusetts 02138, USA}
\author{Mark J.~Bowick}
\email[E-mail:]{mjbowick@syr.edu}
\affiliation{Soft Matter Program and Department of Physics, Syracuse University, Syracuse, New York 13244-1130, USA}
\affiliation{Kavli Institute for Theoretical Physics, University of California, Santa Barbara, California 93106-4030, USA}

\date{\today}
\begin{abstract}
We use molecular dynamics to study the vibrations of a thermally fluctuating two-dimensional elastic membrane clamped at both ends. We directly extract the eigenmodes from resonant peaks in the frequency domain of the time-dependent height and measure the dependence of the corresponding eigenfrequencies on the microscopic bending rigidity of the membrane, taking care also of the subtle role of thermal contraction in generating a tension when the projected area is fixed.  At finite temperatures we show that the effective (macroscopic) bending rigidity tends to a constant as the bare bending rigidity vanishes, consistent with theoretical arguments that the large-scale bending rigidity of the membrane arises from a strong thermal renormalization of the microscopic bending rigidity. Experimental realizations include covalently-bonded two-dimensional atomically thin membranes such as graphene and molybdenum disulfide or soft matter systems such as the spectrin skeleton of red blood cells or diblock copolymers.   
\end{abstract}

\pacs{68.60.Dv, 62.25.Jk, 46.70.-p, 62.20.dq}
\maketitle

\section{Introduction}
Since the discovery of graphene about ten years ago \cite{Novoselov2005}, atomically thin, two-dimensional (2D) crystals such as molybdenum disulfide \cite{Wang2012}, boron nitride \cite{Song2010} and black phosphorus \cite{Li2014} have attracted considerable attention \cite{Geim2013}. While not atomically thin, the spectrin skeleton of red blood cells \cite{Schmidt1993} and diblock copolymer systems \cite{Shum2008} provide alternative arenas in which to investigate the physics of membranes and shells, in this case with thicknesses of order 1-10 \mbox{nm}. Self-consistent statistical field theory treatments of coarse-grained models of polymerized membranes, flexible elastic sheets and shells predict that thermally-driven shape undulations (sometimes called flexural phonons) generate a non-linear stretching term in the effective free energy in addition to the usual bending energy contribution \cite{Nelson2004, Bowick2001, Paulose2012}. The isotropic shape corrugations generated by height fluctuations of these flexible elastic sheets render the dressed bending rigidity strongly scale-dependent, \textit{growing} as a power law $l^{\eta}$, where $l$ is a measure of the spatial extent of the shape fluctuations \cite{Nelson1987,Kantor1987}. The first treatment \cite{Nelson1987} employed a one-loop expansion of the bending energy, within a self-consistent approximation, and gives $\eta=1$. A more elaborate self-consistent theory allowing for the renormalization of the in-plane elastic constants yields $\eta = 0.821$ \cite{Doussal1992}. Other approaches lead to refined values of $\eta$ analytically \cite{Gazit2009,Kownacki2009,Braghin2010,Troster2013,Troster2015} and allow comparisons with simulation results \cite{Bowick1996,Los2009}.

A revealing proxy for the field theory calculation is provided by a standard piece of writing paper. When flat it is very difficult to stretch but is so floppy it cannot even support its own weight. Randomly crumpling the paper and then ironing it out introduces ripples or corrugations running isotropically across the sheet (the analog of thermal fluctuations), which is now far stiffer to bend \cite{Kosmrlj2013} and can indeed support its own weight. Field theory calculations indicate that thermally induced ripples have precisely the same dramatic effect on the bending modulus. Thermally fluctuating flexible elastic sheets are thus more rigid macroscopically than microscopically. As opposed to quenched static ripples resulting from, say, frozen-in grain boundaries arising during sample preparation, the thermally-induced ripples we study are dynamic, which allows us to study their height profile in frequency space. Recent room temperature experiments on micron-length graphene ribbons have carefully measured the macroscopic bending rigidity of the ribbons, finding values up to four orders of magnitude higher than the bare value, though the relative contribution from frozen static ripples and thermal fluctuations remains to be thoroughly explored \cite{Blees2015}. 

A direct determination from computer simulations of the effective bending rigidity of a thermalized ribbon, clamped on two sides as in the experiments \cite{Blees2015}, has, however, been lacking to this point. Here we use molecular dynamics (MD) to simulate the thermally induced shape fluctuations of atomically thin ribbons like graphene. Fourier analysis of the time-dependent height clearly reveals the dominant ribbon eigenmodes. By tracking the dependence of the eigenfrequency of a fixed mode on the bare bending rigidity, and taking care of effects on the eigenfrequency from the tension exerted by clamped boundaries that impose a fixed projected area, we find an effective bending rigidity that stiffens, qualitatively in agreement with the coarse-grained field theory methods outlined above. 

Our method may be used for arbitrary boundary conditions but for simplicity here we clamp both ends of the ribbon across the whole width to avoid macroscopic crumpling at small bare bending rigidity and other complications that distract from the main task of uncovering the fluctuation-dressed effective bending rigidity. Our boundary conditions effectively impose a constraint of constant projected area on the ribbon. They can easily be achieved experimentally via the techniques of Ref.~\cite{Blees2015} by fixing the positions of gold pads that anchor the ends of otherwise freely suspended graphene ribbons in water at room temperature. 

As we demonstrate explicitly with our simulations (see Fig.~\ref{force}), the effect of the clamped boundary conditions is to generate a uniaxial tension that depends on the ratio of $k_{B}T$ to the bare bending rigidity $\kappa$. This tension acts like an ordering field on membrane normals. The effect of an isotropic version of this remarkable entropically generated tension was studied years ago by Guitter et al. \cite{Guitter1989}, who uncovered a ``crumpling transition" associated with constraints that impose a fixed projected area on the flat phase of thermally wrinkled materials. See Refs.~\cite{Kosmrlj2016} and \cite{Gornyi2017} for recent studies of the (nonlinear) response of thermalized atomically-thin free-standing materials to both isotropic and uniaxial tensions. Bonilla and Ruiz-Garcia \cite{Bonilla2016} have used the saddle point method of Ref.~\cite{Guinea2014} to study the critical radius and temperature for buckling in graphene in the large $d$ limit, where $d$ is the embedding or bulk dimension. For a review of the novel effect of strains on graphene and other two-dimensional materials, see the work by Amorim et al. \cite{Amorim2016}. For an experimental study of the mechanical properties of free-standing graphene, see Ref.~\cite{Nicholl2015}.

\section{Simulation method}
In thin plate elasticity theory the free energy of an isotropic thin plate can be described as a sum of stretching and bending terms \cite{LandauVol7, Audoly2010}: $F_{el}=F_{s}+F_{b}$. With the membrane represented as a discrete triangular lattice, the stretching energy is \cite{Seung1988,Lidmar2003,Schmidt2012} 
\begin{equation}
F_{s} = \frac{\varepsilon}{2}\sum_{\left< ij \right>}\left(\left|\mathbf{r}_{i}-\mathbf{r}_{j} \right| -a \right)^{2}
\label{Fs_dis},
\end{equation}
and the bending energy is 
\begin{equation}
F_{b} = \frac{\tilde{\kappa}}{2}\sum_{\left< I J \right>}\left(\mathbf{\hat{n}}_{I}-\mathbf{\hat{n}}_{J}\right)^{2}
\label{Fb_dis},
\end{equation}
where $\varepsilon$ is the discrete spring constant, $a$ is the equilibrium spring length and $\tilde{\kappa}$ is the discrete bending modulus. As usual $\left<ij\right>$ denotes pairs of nearest-neighbor vertices, with positions $\mathbf{r}_{i}$ in 3D Euclidean embedding space and $\left<IJ\right>$ denotes pairs of triangular plaquettes sharing a common edge and $\mathbf{\hat{n}}_{I}$ are their unit normals. The corresponding continuum moduli are $Y= 2\varepsilon/\sqrt{3}$, $\kappa = \sqrt{3}\tilde{\kappa}/2$ and Poisson ratio $\nu = 1/3$ \cite{Seung1988,Lidmar2003,Schmidt2012}. $a$ is a lattice constant which we can take as the distance between neighboring hexagons if we wish to model a honeycomb structure like graphene. The parameters $\varepsilon$ and $\tilde{\kappa}$ can be adjusted to match the unrenormalized elastic parameters of, e.g., graphene, whose some reported Poisson ratio values are in the range of 0.14 to 0.46 (see Refs.~\cite{Reddy2006, Cao2014} and references therein), not too far from the value $\nu = 1/3$. This discretized model has been used to study a wide variety of 2D elastic membranes such as viral capsids \cite{Lidmar2003,Nguyen2005,Siber2006}, graphene \cite{Zhang2014}, pollen grains \cite{Katifori2010}, elastic shells \cite{Wan2015} and capsules \cite{Vliegenthart2011}. Here we apply this model to investigate the statistical mechanics of sheets of thin 2D membranes, with applications to both atomically thin \cite{Geim2013} and slightly thicker materials \cite{Schmidt1993,Shum2008}, and to gain an understanding of the effect of thermal fluctuations on the mechanical properties of these membranes with clamped boundary conditions as our primary goal. With graphene as a concrete example, we set the equilibrium spring length $a$ to be $\sqrt{3}a_0$, where $a_0=1.42 \, \mbox{\AA}$ is the carbon-carbon bond length in graphene {--} thus $a \approx 2.46 \, \mbox{\AA}$. The triangular lattice we employ is the dual of graphene's honeycomb lattice. To give the correct graphene density we take the mass of every vertex to be $m = 2 m_{C} \approx 4 \times 10^{-26}$ kg, where $m_{C}$ is the mass of a carbon atom. In Ref.~\cite{Bowick2016}, it was shown that this simplified dual model gives results consistent with graphene simulations \cite{Zakharchenko2010} that incorporate explicitly the energetics of carbon-carbon bonds. We choose $a$ and $m$ as our units of length and mass and set $a=1$ and $m=1$ in all simulations. Figure~\ref{height}(a) displays the initial, zero-temperature flat configuration of the membrane in the $x-y$ plane, with $n_{1}=32$ vertices in the long direction and $n_{2}=11$ vertices staggered along the short direction ($L_{0}\approx 30a\approx 74\mbox{\AA}$ and $W_{0}=5\sqrt{3}a \approx 21\mbox{\AA}$). There are $347$ vertices in total. We clamp the edge vertices along the two zigzag boundaries indicated by the orange lines in Fig.~\ref{height} and tag a marked vertex (shown in red). We perform MD simulations using the HOOMD-blue \cite{Hoomdblue, Anderson2008, Glaser2015} and the LAMMPS software packages \cite{Plimpton1995}, both giving consistent results. We choose $E_{0}= 1 \, \mbox{eV}$ as the unit of energy and $k_{B}T=0.025 \, \mbox{eV}$ (corresponding to room temperature) in all simulations. The elastic free energy is calculated as the sum of the stretching and bending energies given in Eqs.~(\ref{Fs_dis}) and (\ref{Fb_dis}), with unrenormalized elastic parameters $\kappa =1.2 \, \mbox{eV}\, = 48 \, k_{B}T$ \cite{Nicklow1972,Fasolino2008} and $Y=20 \, \mbox{eV}/\mbox{\AA}^{2} = 800 \, k_{B}T/\mbox{\AA}^{2}$ \cite{Lee2008,Zhao2009} as parameter estimations for graphene. The discrete parameters $\varepsilon$ and $\tilde{\kappa}$ follow from the relations above. After giving the free vertices a small random out-of-plane displacement, we update their positions in the constant temperature (NVT) ensemble. The simulation unit of time thus corresponds to a real time $t_{0} = m \,a^{2}/E_{0} \approx 0.12 \mbox{ps}$. Finally we set the integration time step to be 0.005. Every simulation timestep $\tau$ thus corresponds to a real time $\tau =0.005 \, t_{0} \approx 0.6 \, \mbox{fs}$ (movie 1).

\section{Results and discussion}
\begin{figure}
\centering 
\includegraphics[width=3.2in]{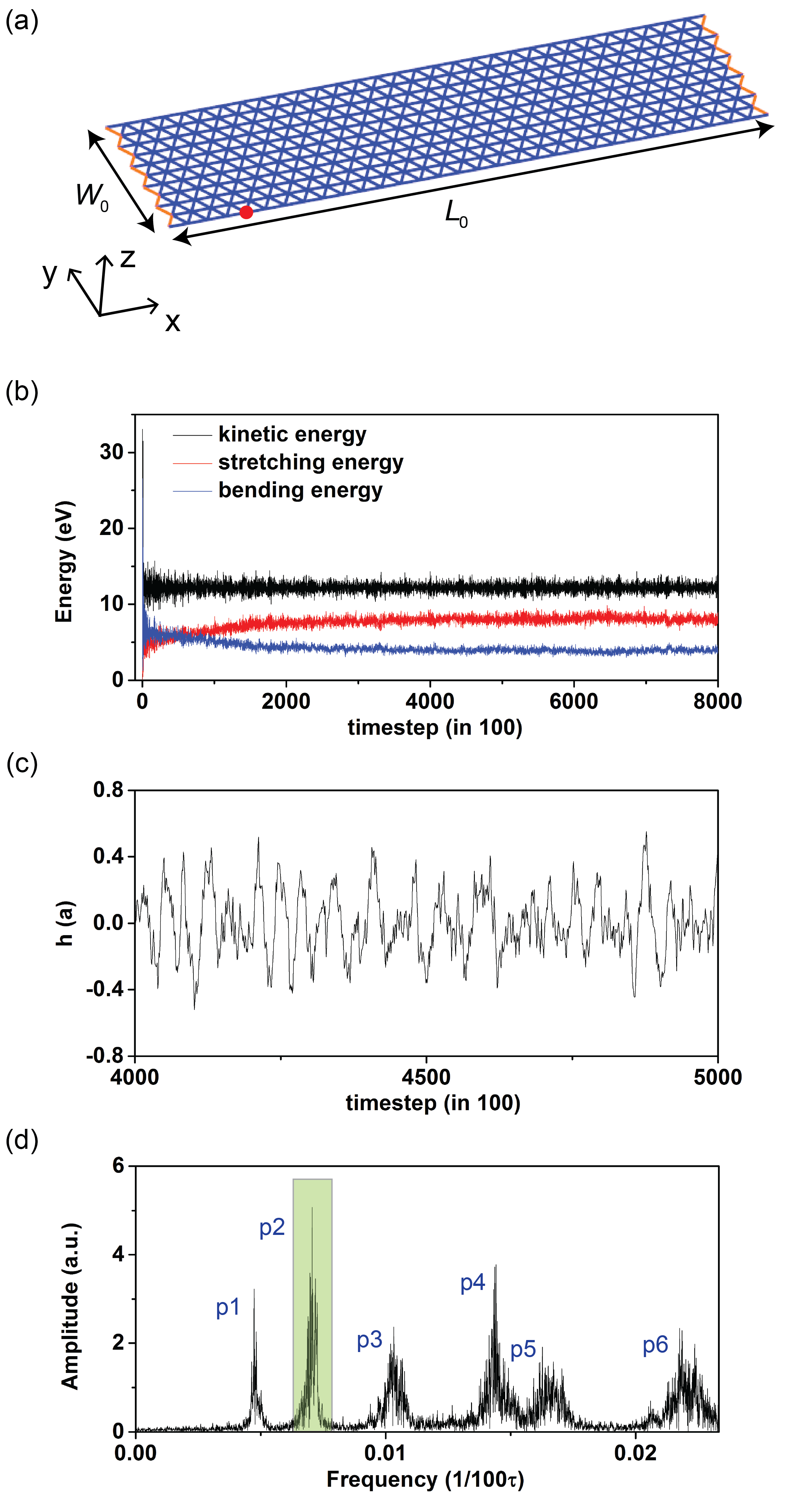}
\caption{(Color online) (a) A triangulated membrane with $n_{1}=32$ vertices along the length and $n_{2}=11$ vertices, in staggered form, along the width. The vertices in the two zigzag boundaries are clamped (orange lines). We label a particular vertex on the front edge with a large red dot. The snapshot was generated using the Visual Molecular Dynamics (VMD) package \cite{Humphrey1996} and rendered using the Tachyon ray tracer \cite{Stone1998}. (b) Energies of the system during the first $8\times 10^{5}$ timesteps. (c) Height of the red vertex as a function of time after equilibrating for $4\times 10^{5}$ time steps. (d) The Fourier amplitudes of the first six peaks in the frequency domain of the height function of the red vertex over $10^7$ timesteps. The light green shaded rectangle near p2 is explained in the text.}
\label{height}
\end{figure}

\begin{figure}
\centering 
\includegraphics[width=3.2in]{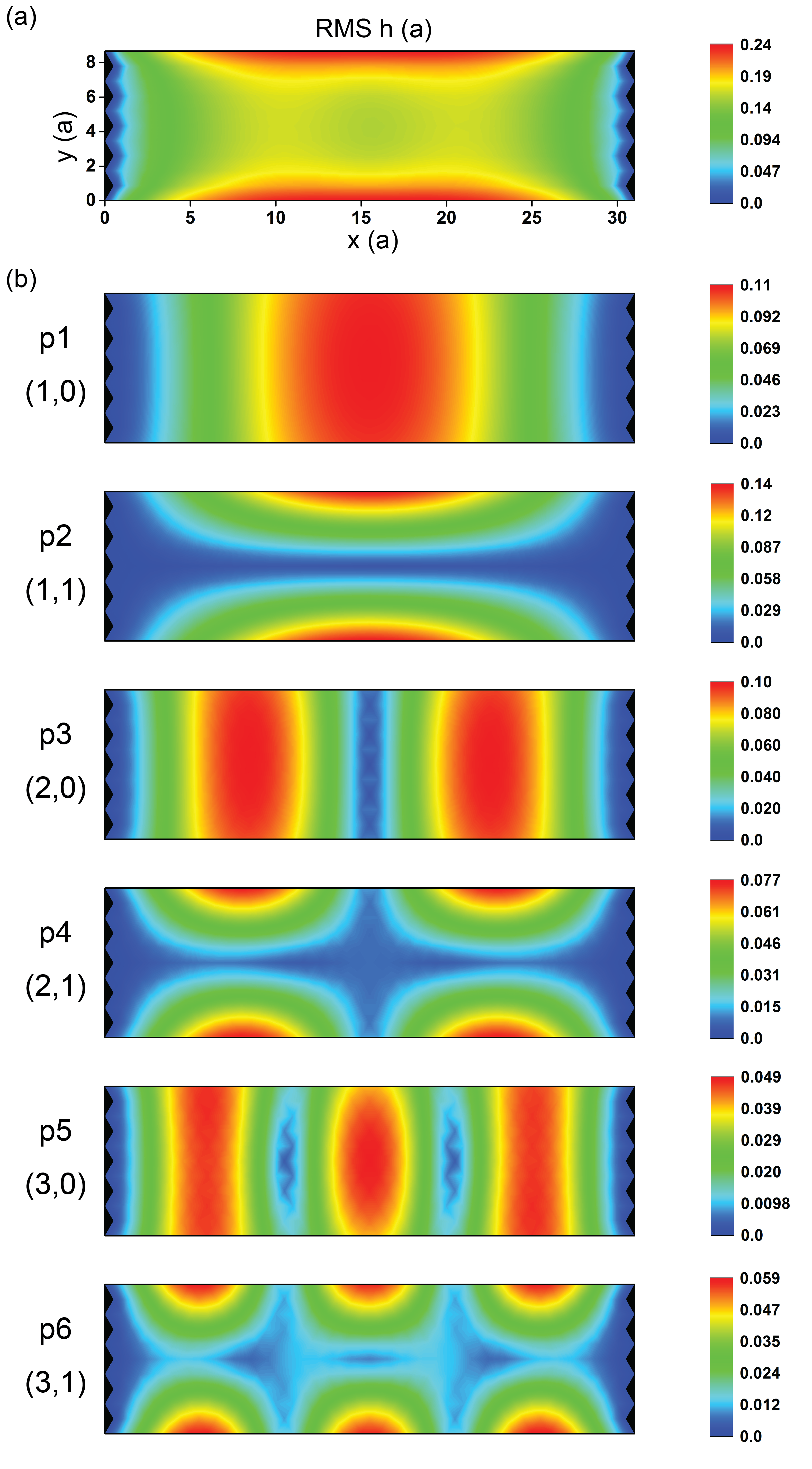} 
\caption{(Color online) (a) The root mean square height $h_{\scriptsize\mbox{rms}}$ of the height profile (e.g.,~Fig.~\ref{height}(c) for the labeled vertex) as a function of position on the ribbon. (b) The corresponding $h_{\scriptsize\mbox{rms}}$ for the first six modes in Fig.~\ref{height}(d).}
\label{modes}
\end{figure}

\begin{figure}
\centering 
\includegraphics[width=3.3in]{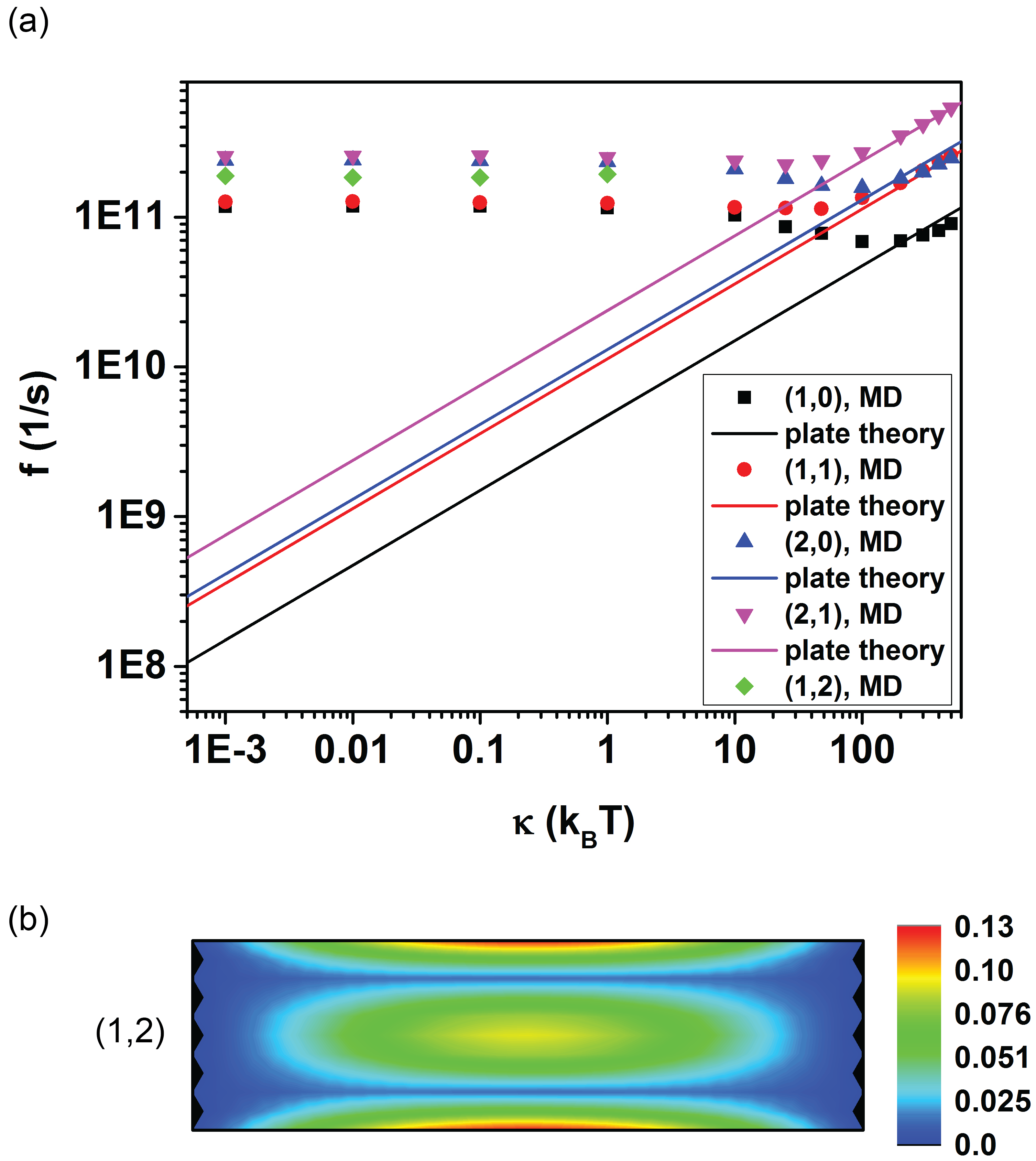} 
\caption{(Color online) (a) Log-log plot of the frequencies of modes for various $\kappa$ values. Lines are the predictions of the thin plate theory. (b) The configuration of the (1,2) mode that appears in the small $\kappa$ range. } 
\label{frequency}
\end{figure}

We log a sample configuration every $100$ simulation timesteps. Figure \ref{height}(b) shows the kinetic, stretching and bending energy profiles for the first $8\times 10^{5}$ timesteps in a run of $10^{7}$ timesteps in all. The ribbon equilibrates after roughly $3\times 10^{5}$ timesteps. Figure \ref{height}(c) displays the time-dependent height ($z$-coordinate) of a tagged vertex (the red vertex in Fig.~\ref{height}(a)) for $10^5$ timesteps after equilibrium. The vertex fluctuates about the $x-y$ plane ($z=0$). To get a better understanding of the height fluctuations, we Fourier transform the height profile of Fig.~\ref{height}(c) and plot the amplitude of the first six peaks in frequency space in Fig.~\ref{height}(d). The root mean square height of the membrane (Fig.~\ref{modes}(a)) has contributions from all the excited modes. To extract the contribution from a given mode we select the appropriate peak in the frequency domain by applying a window function as illustrated by the light green shading centered on peak 2 (p2) in Fig.~\ref{height}(d). We then inverse Fourier transform to extract the time profile of the height for this mode. The same window filter is used for all vertices. Figure~\ref{modes}(b) displays the first six modes. As stiff, atomically thin materials like graphene bend more readily than they stretch, the in-plane displacements of a vertex are small compared to the out-of-plane displacement, and we can neglect variations of the $x$ and $y$ coordinates of vertices when plotting Fig.~\ref{modes}. The frequency of a peak is determined as the center of the best fit to a Gaussian. On the other hand, the ribbon is well-approximated by a thin elastic plate \cite{Chowdhury2011,Chandra2012,Liu2012}. 
We calculate the frequencies of eigenmodes of a rectangular plate of dimension $L_{0}\times W_{0}$, with clamped-clamped boundary conditions along the edges of length $W_{0}$, free-free boundary conditions along the remaining edges, using the classical plate theory \cite{Blevins1984, Leissa1969}. We find that the lowest six modes are, in order of ascending frequency, $(1,0)$, $(1,1)$, $(2,0)$, $(2,1)$, $(3,0)$ and $(3,1)$ ( $i$ and $j$ in the mode index $(i,j)$ are equal to the number of crossing points with the $x-y$ plane in the $x$ direction (clamped boundaries not included) plus 1, and the number of crossing points in the $y$ direction, respectively), matching the height maps in Fig.~\ref{modes}.
We now explore the dependence of the eigenfrequencies on the bare bending rigidity $\kappa$ (keeping $Y$ fixed), thus enhancing the importance of out of plane thermal fluctuations since we keep temperature fixed at $k_{B}T=0.025 \, \mbox{eV}$ (room temperature). Figure \ref{frequency}(a) plots the eigenfrequencies of the lowest four modes as a function of $\kappa$. The straight lines are predictions of the classical plate theory (assuming straight boundary conditions and no pre-tension), where the frequency $f \sim \kappa^{1/2}$. When $\kappa$ is large ($\kappa \geq 100k_{B}T$), the frequencies of the eigenmodes are close to those expected by the classical plate theory; when $\kappa$ is small ($\kappa \leq 1 k_{B}T$), however, the frequencies can be orders of magnitude higher than their predicted values. Indeed, the eigenfrequencies tend to a constant as the bare $\kappa$ vanishes. Note also that an additional mode (1,2), shown in Fig.~\ref{frequency}(b), which is a high frequency mode predicted by the plate theory (not among the six modes in Fig.~\ref{modes}), joins the low frequency modes of classical plate theory in the small $\kappa$ range.


The increase of these resonant frequencies has two origins. The first is thermal contraction. At zero temperature (without any initial perturbation), the ribbon is flat; at finite temperature, the ribbon tends to contract due to entropy \cite{Guitter1988,Guitter1989}. In other words, the projected area of a ribbon at finite temperature (if the clamps at the ends of the longer $L_{0}$ direction are allowed to slide) is smaller than that of the flat configuration at $T=0$ \cite{Liang2016}. When we clamp the two opposite ends at, e.g., the length of the zero-temperature flat configuration $L_{0}$, the fixed boundaries are effectively pulling on the ribbon as it vibrates which raises the eigenfrequencies. The second cause is the stiffening effect of shape undulations resulting from thermal fluctuations, as outlined in the Introduction. 

\begin{figure}
\centering 
\includegraphics[width=3.4in]{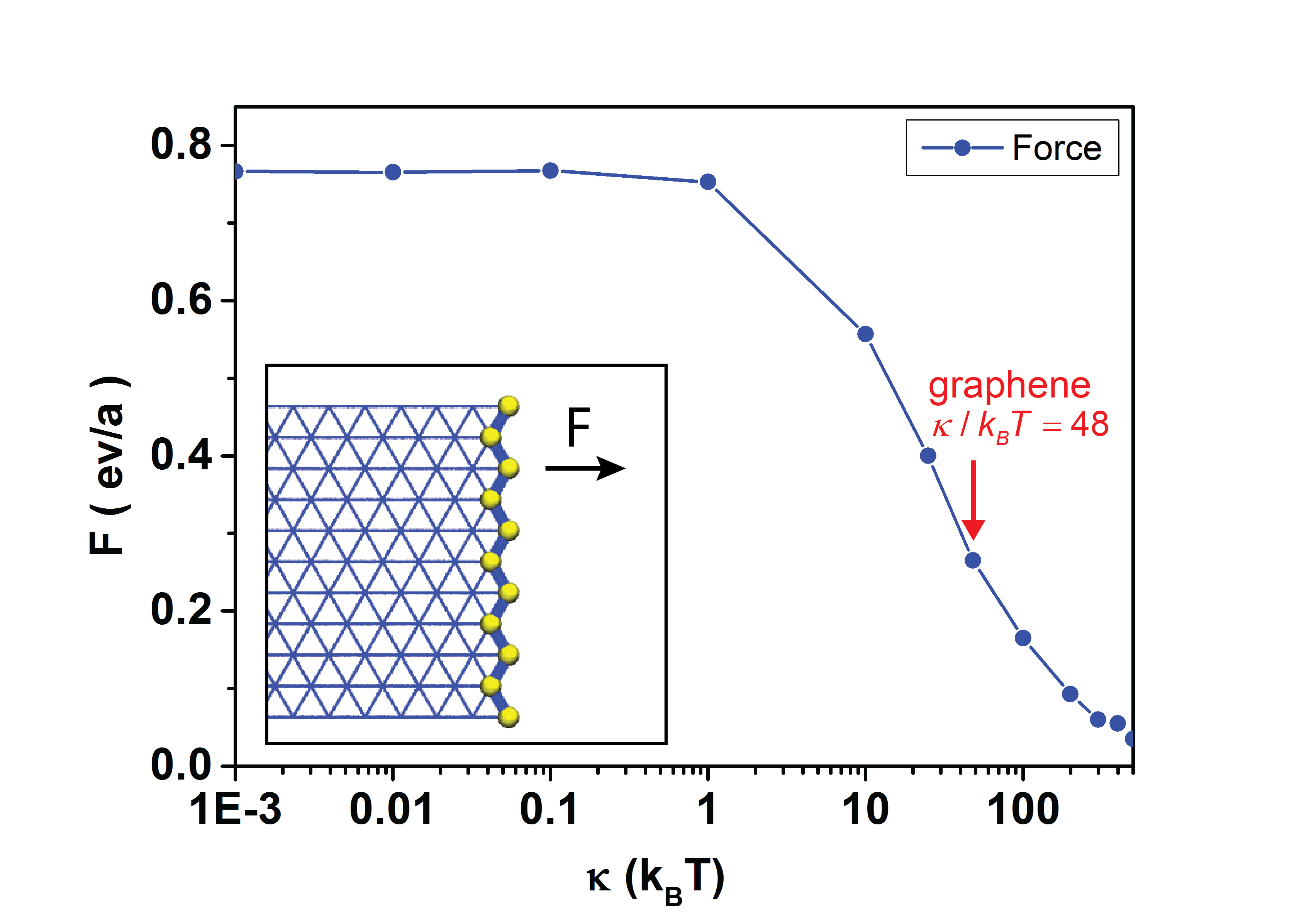} 
\caption{(Color online) Force $F$ on each vertex at the sliding end needed to pull the ribbon back to the flat configuration. Inset: The sliding end with an external force applied to a zigzag array of vertices at the right edge. Vertices in the sliding end (yellow) are constrained to move in the $x$ direction and connected by edges with a high spring constant (thick blue lines). The red arrow indicates the $\kappa$ value of graphene.} 
\label{force}
\end{figure}

To account for the effect of thermal contractions in the presence of clamping, and isolate contributions due solely for thermal ripples, we measure the tension exerted by the boundaries. For this purpose it is easier to study a ribbon with one clamped end and one sliding end. The sliding end is constructed as shown in the inset of Fig.~\ref{force}. We constrain the 11 vertices in the zigzag boundary to move in $x$ direction (using the lineforce command in LAMMPS) and choose the edges connecting these vertices to have a high spring constant. These vertices will then move together horizontally. We have taken into consideration the corresponding subtraction of the total number of degrees of freedom in the simulation. When $\kappa$ is large the sliding end fluctuates and the amplitude of fluctuation is small (movie 2). As $\kappa$ decreases, the amplitude of fluctuations gets larger until $\kappa \lesssim 1k_{B}T$, where the ribbon crumples (movie 3 and 4). We then add a horizontal force on the sliding end. We increase the force gradually until the ribbon is pulled back to have projected length $L_{p}$ equal to its zero-temperature length $L_{0}$. The crumpled ribbon becomes flat when the force is large enough (movie 5). Figure \ref{force} plots the measured force $F$ on each vertex at the sliding end as a function of $\kappa$. When $\kappa$ is large the force is small. As $\kappa$ decreases the force increases until the small $\kappa$ range, where the force tends to a constant. The response of thermalized ribbons to this type of pulling (as well as bending) has been analyzed in Ref.~\cite{Kosmrlj2016} and the non-Hookean statistical mechanics of thermalized ribbons clamped at one end only, with the associated scale-dependent stiffening of the bending rigidity, studied in Ref.~\cite{Bowick2016}.

\begin{figure}
\centering 
\includegraphics[width=3.2in]{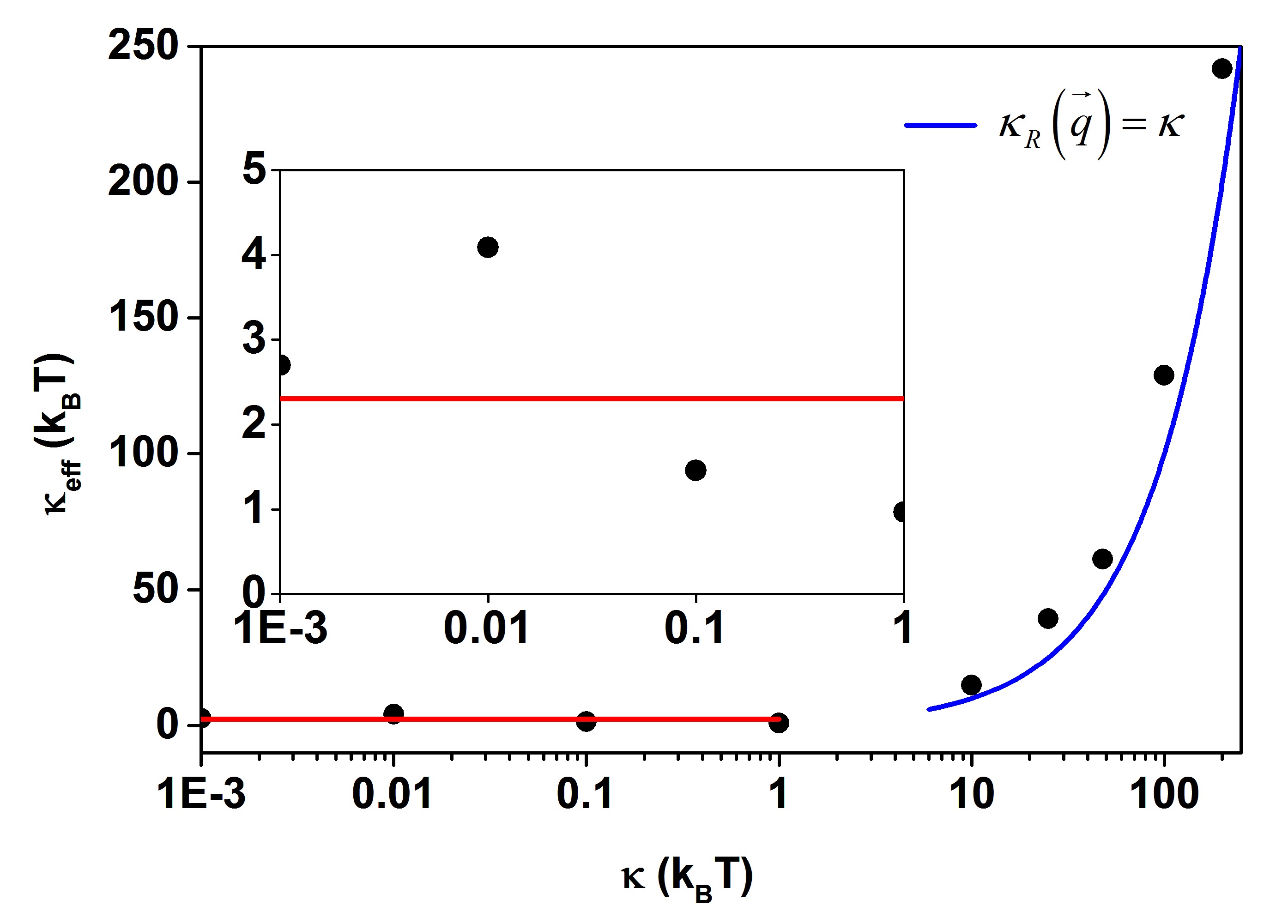} 
\caption{(Color online) The effective bending rigidity $\kappa_{\scriptsize\mbox{eff}}$ of the (1,1) mode as a function of the bare bending rigidity $\kappa$. Red line: Average of the four points in the small $\kappa$ range. Inset: Zoom in of the small $\kappa$ range.} 
\label{kappa_eff}
\end{figure}

We then use COMSOL to solve the vibration of a plate subjected to an in-plane tension along the long direction. We use the 2D Plate modulus in COMSOL to simulate the ribbon subject to an in-plane stress given by $11F/W_{0}$, corresponding to a force on each of the 11 vertices on the right side of the ribbon (inset to Fig.~\ref{force}). We sweep over the bending rigidity $\kappa_{C}$ ($C$ is short for COMSOL, to distinguish it from the bending rigidity in MD simulations) of the ribbon, until the frequency of a certain mode matches that of the MD simulation. We call this $\kappa_{C}$ the effective bending rigidity $\kappa_{\scriptsize\mbox{eff}}$. As an example, Fig.~\ref{kappa_eff} plots $\kappa_{\scriptsize\mbox{eff}}$ of the $(1,1)$ mode as a function of $\kappa$. In the small $\kappa$ range $\kappa_{\scriptsize\mbox{eff}}$ tends to be constant and is of order several $k_{B}T$ (see inset of Fig~\ref{kappa_eff}). On the other hand, the self-consistent theory of elastic membranes implies that non-linear stretching results in a renormalized bending rigidity $\kappa_{R}$ determined by the integral equation \cite{Nelson1987}
\begin{widetext}
\begin{eqnarray}
\kappa_{R}(\vec{q}) = \kappa  +  k_{B}T Y \int \frac{d^{2}k }{\left( 2 \pi \right)^{2}} \frac{ \left[ \hat{q}_{i} P_{ij}^{T}\left(\vec{k}\right) \hat{q}_{j}  \right]^{2} } { \kappa_{R}(\vec{q}+\vec{k}) | \vec{q}+ \vec{k}|^{4}+\sigma_{ij}(\vec{q}+ \vec{k})_{i} (\vec{q}+ \vec{k})_{j}},
\label{kappa_renormalized}
\end{eqnarray} 
\end{widetext}
where $\vec{q}$ is the wave vector, $\hat{q}$ is the corresponding unit vector, $P_{ij}^{T}\left(\vec{k}\right)=\delta_{ij}-k_{i}k_{j}/k^{2}$ is the transverse projection operator and $\sigma_{ij}$ is an external edge tension \cite{Kosmrlj2016}\cite{Roldan2011}. In our case $\sigma_{xx}=11F/W_{0}$ and other components of $\sigma_{ij}$ vanish. When $\kappa$ is large, the correction term is small compared to the bare $\kappa$ and  thus $\kappa_{R}(\vec{q})\approx\kappa$. In the small $\kappa$ limit, as the external force tends to a constant (Fig.~\ref{force}), we expect from Eq.~(\ref{kappa_renormalized}) that $\kappa_{R}(\vec{q})$ will also tend to a constant. We now estimate this limiting value. Taking, as the lowest order approximation, $\kappa_{R}(\vec{q}+\vec{k})\approx \kappa_{R}(\vec{q})$  on the right hand side of Eq.~(\ref{kappa_renormalized}) and approximating the $(1,1)$ mode shape by $\vec{q}=(\pi/L_{0}, \pi/W_{0})$ , with the upper and lower limits for $\vec{k}$ being $k_{max}=\pi/a$ and $k_{min}=\pi/L_{0}$, we solve Eq.~(\ref{kappa_renormalized}) numerically and obtain $\kappa_{R}(\vec{q})\approx 21 k_{B}T$. With $k_{max}=\pi/a$ and $k_{min}=\pi/W_{0}$, $\kappa_{R}(\vec{q})\approx 13 k_{B}T$. Thus $\kappa_{R}(\vec{q})$ is consistent with our determination of the effective bending rigidity. There are, however, several possible sources of discrepancy. There may be higher order corrections to $\kappa_{R}(\vec{q})$. In addition, the frequency given by COMSOl is only accurate for small amplitudes, whereas we necessarily have finite amplitudes. There may also be corrections due to the scale dependence of the Young's modulus and the Poisson ratio which we neglect. Finally, there can be finite-size effects associated with the boundary conditions.  

The order of the modes also puts constraints on $\kappa_{\scriptsize\mbox{eff}}$. In the MD simulation the frequencies and shapes of the modes and the forces needed to pull the ribbon back to a projected length $L_{p}=L_{0}$, all approach limiting values, in the small $\kappa$ limit. When we sweep over $\kappa_{C}$ to match the frequency of a certain mode in COMSOL, in contrast, we find that the order of modes changes with $\kappa_{C}$. However, at $\kappa_{C} = \kappa_{\scriptsize\mbox{eff}}$ not only the frequency matches but also the order of modes matches. More exactly, for the small $\kappa$ range ($\kappa \le 1 k_{B}T$), if we only want the order of the modes in COMSOL to match that in MD simulation, then $\kappa_{\scriptsize\mbox{eff}}$ is confined to be about $2\sim 12 k_{B}T$ (see e.g., $\kappa_{C}=5 k_{B}T$ in Table.~\ref{order}). If there is no renormalization, the eigenmodes for small $\kappa_{C}$ values at this relatively large stress ($F=0.768 \, \mbox{ev}/a$ corresponding to the force at $\kappa=0.1 k_{B}T$ in the MD simulation ), exhibit wrinkles only along the unstretched direction (e.g., the (1,4), (1,5) modes at $\kappa_{C}=0.1 k_{B}T$ in Table.~\ref{order}). These modes are similar to the profile of an elastic membrane stretched along one direction (see, e.g., Ref.~\cite{Takei2011}). Thus the order of the modes also supports the presence of a strong thermal renormalization.     
\begin{table}
\centering
\caption{Lowest six modes ordered by frequency when sweeping over $\kappa_{C}$ in COMSOL, with the stress $11F/W_{0}$ and $F=0.768 \, \mbox{ev}/a$ fixed.}
\begin{tabular}{L{1.5cm}C{1cm}C{1cm}C{1cm}C{1cm}C{1cm}C{1cm}}
\hline 
\hline
$\kappa_{C}$ ($k_{B}T$) & 1 & 2 & 3 & 4 & 5 & 6\\
\hline
0.1          & (1,2) & (1,3) & (1,4) & (1,5) & (2,2) & (2,3)\\
1            & (1,0) & (1,1) & (1,2) & (1,3) & (2,0) & (2,1)\\
5            & (1,0) & (1,1) & (1,2) & (2,0) & (2,1) & (2,2)\\
14.4         & (1,0) & (1,1) & (2,0) & (1,2) & (2,1) & (2,2)\\
20.8         & (1,0) & (1,1) & (2,0) & (2,1) & (1,2) & (3,0)\\
\hline
\end{tabular}
\label{order}
\end{table}

\section{Conclusions}
In summary, we have used molecular dynamics and a simple coarse-grained model to simulate atomically thin 2D membranes and study their thermally-induced fluctuations. We have identified the eigenmodes of the system and find that thermal effects significantly change the eigenfrequencies predicted by classical plate theory, qualitatively consistent with the scale-dependent renormalized bending rigidity predicted by the statistical field theory of elastic membranes.    

\acknowledgments  
We thank Rastko Sknepnek, Teng Zhang and Meng Xiao for helpful discussions. We thank the Syracuse University campus OrangeGrid (NSF award No. ACI-1341006) for computational resources. The work of MJB and DRN was supported by the National Science Foundation through the NSF DMREF program via grants DMR-1435794 and DMR-1435999. The work of MJB and DW was also supported by the Syracuse Soft Matter Program. Finally MJB was supported in part by the NSF under Grant No. NSF PHY11-25915.\\

\appendix
\section{Movies}
Movie 1 shows the vibration of a two-end clamped ribbon with $\kappa= 48 \, k_{B}T$ over a period of time after reaching equilibrium. The configuration is logged every $500\tau$. Movies 2 to 4 show the evolution of a ribbon with one-clamped end and the opposite end sliding, and with $\kappa=48, 1, 0.001 \, k_{B}T$, respectively, starting from the initial perturbed configuration. As $\kappa$ decreases, the sliding end first fluctuates with a small amplitude, then with a larger amplitude, and finally crumples. The frequency of logging is $500\tau$, $600\tau$, $500\tau$, respectively. Movies 5 has the same setup as that in movie 4, but with a horizontal force $F=0.8 \, \mbox{ev}/a$ on each vertex at the sliding end. The ribbon remains flat over the simulation time. In all movies, the geometric parameters and the temperature are the same as those in the main text. The clamped ends are colored yellow and the sliding ends blue. Movies can be found in Ref.~\cite{movies}.  

\bibliographystyle{apsrev4-1}
\bibliography{ribbon_references}

\end{document}